# Machine Learning Quantum Reaction Rate Constants

Evan Komp[1] and Stéphanie Valleau,[1,*]

[1]*Department of Chemical Engineering, University of Washington, Seattle, Washington 98195, United States*

*corresponding author: valleau@uw.edu

**ABSTRACT**

The ab initio calculation of exact quantum reaction rate constants comes at a high cost due to the required dynamics of reactants on multidimensional potential energy surfaces. In turn, this impedes the rapid design of the kinetics for large sets of coupled reactions. In an effort to overcome this hurdle, a deep neural network (DNN) was trained to predict the logarithm of quantum reaction rate constants multiplied by their reactant partition function – rate products. The training dataset was generated in-house and contains ~1.5 million quantum reaction rate constants for single, double, symmetric and asymmetric one-dimensional potentials computed over a broad range of reactant masses and temperatures. The DNN was able to predict the logarithm of the rate product with a relative error of 1.1%. Further, when comparing the difference between the DNN prediction and classical transition state theory at temperatures below 300K a relative percent error of 31% was found with respect to the exact difference. Systems beyond the test set were also studied, these included the H + $H_2$ reaction, the diffusion of hydrogen on Ni(100), the Menshutkin reaction of pyridine with $CH_3Br$ in the gas phase, the reaction of formalcyanohydrin with $HS^-$ in water and the F + HCl reaction. For these reactions, the DNN predictions were accurate at high temperatures and in good agreement with the exact rates at lower temperatures. This work shows that one can take advantage of a DNN to gain insight on reactivity in the quantum regime.

**INTRODUCTION**

Understanding, evaluating and predicting chemical reactivity computationally is of great interest as it is critical for the design of new materials,[1,2] to tune the production of compounds in industry,[3] for drug design[4] and so forth. For the last century, theories and their application to the calculation of reaction rate constants have been developed and explored extensively.[5–7] Classical, semi-classical and quantum mechanical approaches all require some knowledge of the potential energy surface, ranging from an activation energy, to a minimum energy path, to a full potential energy surface. Classical static theories such as transition state theory (TST) or variational transition state theory[8] are frequently employed due to their low computational cost. However, they come with the need to find a minimum energy path ab-initio and are limited by being valid mostly at high temperatures where quantum effects such as tunneling are less significant. At lower temperatures[9] quantum corrections[10] or approaches such as the Quantum Instanton,[11,12] the flux-flux or flux-side correlation functions take tunneling into account.[13] Unfortunately, fully quantum approaches cannot be applied to systems with many degrees of freedom due to the exponential growth in computational cost with system size.

To overcome these bottlenecks, in the last decade, there has been much interest in using AI machine learning types of approaches to predict chemical properties[14–21] and more recently reactivity.[22–28] The prediction of reaction rate constants using supervised machine learning comes with the requirement of large datasets of reaction rate constants. While some datasets exist for gas phase reactions, e.g. Reaxys[29] or Scifinder[30], the data is largely inhomogeneous and often does not include reaction rate constants. Nonetheless, these datasets have been used to train supervised machine learning algorithms to predict e.g. suitable temperatures for reactions.[23] Recent work[24,27] has also shown promise for the AI prediction of activation energies.

Bowman et al.[26] applied gaussian process regression to a small data set of single barrier reactions and were able to obtain good estimates of the reaction rate constants computed using multidimensional time-dependent hartree (MCTDH), the Quantum Instanton or ring polymer molecular dynamics (RPMD).

In this work we trained and validated a supervised multi-layer perceptron (MLP) deep neural network (DNN) to predict the product of the exact quantum reaction rate constant with the reactant partition function for one-dimensional barriers. The motivation for choosing one-dimensional paths was twofold; on one hand many theoretical methods approximate the overall kinetics with a single most traveled one-dimensional reactive pathway, on the other, given our need for a large dataset, the calculation of rate constants on multidimensional surfaces was unfeasible. Single, double, symmetric and asymmetric barriers were considered. The presence of resonant tunneling for double barrier potentials increases the difficulty of computing the quantum reaction rate constant accurately due to the need to resolve resonant peaks. In this context, a machine learning alternative is more appealing. As shown in Figure 1, we first generated a database of over six million reaction rates by numerical integration of the transmission probability[9] for a broad range of reactant masses, barrier heights, barrier widths, barrier distances, barrier shapes and temperatures. This dataset was then employed to find the optimal hyperparameters for a DNN via grid search, to train the DNN, and ultimately, to predict quantum reaction rate constants.

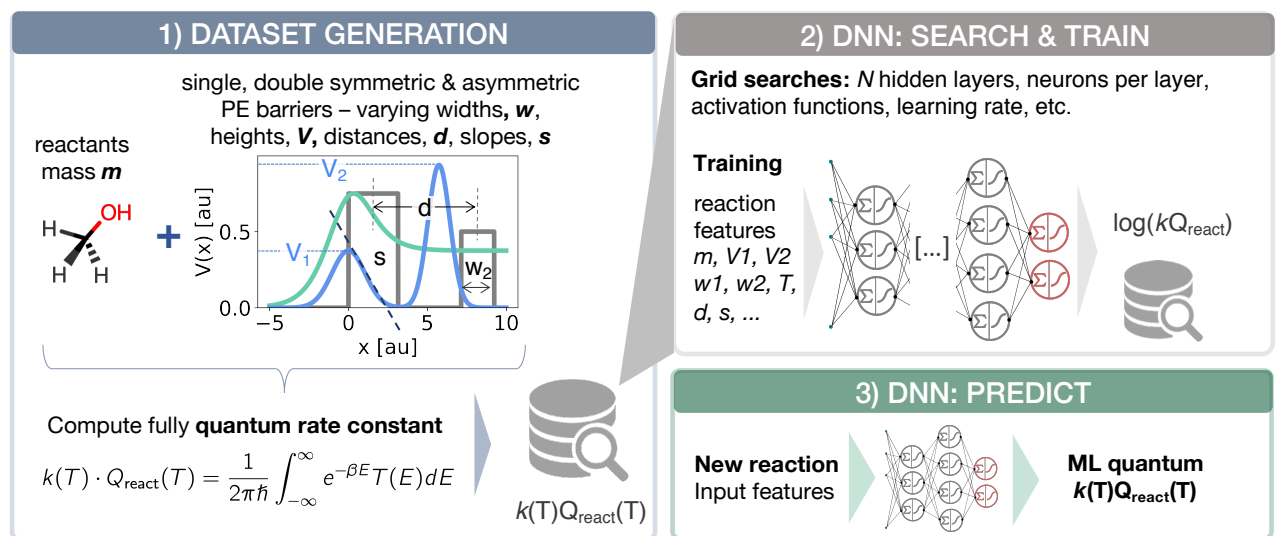

Figure 1: Flowchart of the work carried out in this paper. In panel 1) we show how the dataset was generated. In panel 2) we indicate how the model was constructed by finding optimal hyperparameters with a set of grid searches. Finally, panel 3) shows how the trained multi-layer perceptron deep neural network was employed to predict quantum reaction rate constants.

## COMPUTATIONAL METHODS

### Database Generation

A database of ~6.9 million quantum reaction rate constant products, $k(T)Q_{react}(T)$, was generated for one-dimensional potentials. The quantum reaction rate constant products were computed in the energy domain from the general equation (1)

$$k_Q(T) \equiv k(T)Q_{react}(T) = \frac{1}{2\pi\hbar} \int_{-\infty}^{\infty} e^{-\beta E} T(E) dE \quad (1)$$

where $T(E)$ is the transmission coefficient and $\beta = 1/k_B T$.

The transmission coefficients were evaluated following the approach developed in Ref. [9] where it was shown that given a coordinate position, $x^{(1)}$, located far from the reaction barrier, one can write

$$T(E) = 4 \left| \psi_E(x^{(1)}) + \frac{\hbar}{ip} \partial_x \psi_E(x^{(1)}) \right|^{-2}. \quad (2)$$

In equation (2) the coordinate position, $x^{(1)}$, must satisfy the condition $x^{(1)} \gg L$, where $L$ defines the range $x \in [-L, +L]$ within which the potential is non zero. $\psi_E(x)$ is the eigenfunction of the Hamiltonian for the energy eigenvalue $E = p^2/2m$ where $p$ is the linear momentum and $m$ the reactant mass. This approach was successful in reproducing the full quantum reaction rate constant as could be obtained in the time domain using the flux-flux correlation function approach,[31] with the advantage of avoiding the demanding time propagation of the wavefunction. The reactant partition functions were considered as part of the output to be predicted using machine learning. For simplicity, in this paper we will refer to the output product $k_Q(T) \equiv k(T)Q_{react}(T)$ as the reaction rate constant product. Equation (1) above was solved for single and double symmetric and asymmetric rectangular, Eckart,[32] gaussian and Peskin[33] potentials as shown in panel 1) of Figure 1 and defined below in Table 1.

Table 1: Range of potential energy parameters for the symmetric and asymmetric single and double barriers used to compute reaction rate constants for the dataset. For a symmetric single barrier, $\alpha_1 = 0$. For a symmetric double barrier, $\alpha_2 = 0$.

| Feature | Values |
| --- | --- |
| Reactant mass – $m$ | 1060 – 1060000 [au] |
| Barrier heights – $V_1$, $V_2$ | 5.0 – 80.7 [kcal / mol] |
| Barrier widths – $w_1$, $w_2$ | 0.1 – 6 [au] |
| Distance – $d$ | 0.0 – 5 [au] |
| Asymmetry – $\alpha_1$ | -0.122 – 2.444 |
| Double barrier asymmetry – $\alpha_2$ | -1.832 – 1.832 |
| Temperature – $T$ | 80 – 2500 [K] |

For single rectangular and Eckart potentials exact analytical expressions were used for the transmission coefficient.[7,32,34] Equation (2) is only valid when the energy of reactants is identical to that of the products, however we accounted for single asymmetric barriers by using the analytical expression of the transmission coefficient of asymmetric Eckart potentials. More details on the numerical evaluation of the transmission coefficient can be found in the supporting information.

**Input features and datasets**

A dataset of ~1.5 million points was extracted from the original database to train a deep neural network to predict the reaction rate constant products (eq. 1). To obtain the new dataset, some small mass values were randomly dropped so as to avoid having an unbalanced dataset. Plots of the dataset distributions with respect to input features are shown in the supporting information. Of the new dataset, 80% was used to train and 20% to test the DNN.

The input features (Table 1) included reactant mass, reaction temperature, first and second barrier height, first and second barrier widths, distance between barriers, barrier slope and asymmetry for single and double barriers.

The input features of both the training and testing datasets were normalized with respect to the mean and standard deviation of the training dataset. The output label was chosen to be the logarithm of the rate constant product, $\log k_Q(T)$, to avoid the problem of values ranging over hundreds of orders of magnitude. The output label was then scaled to the range $[-1, 1]$ by using the training dataset label minimum and maximum values.

**Hyperparameters optimization**

The network's optimal architecture and modeling hyperparameters were identified by carrying out grid searches with five-fold cross validation using Scikit-learn's GridSearchCV[35] and Keras with TensorFlow backend.[36] The first search was conducted over one to three hidden layers of $n \in [1, 6, 12, 24, 32, 64, 128]$ neurons per layer with batch size 32, 64 or 128 at 300 epochs. For this grid search, we chose the Adam optimizer[37] with learning rate set to 0.001, $\beta_1 = 0.9$ and $\beta_2 = 0.999$, and the sigmoid and tanh activation functions for the hidden and output layer (See Table S2 for details on the first grid search). Subsequently a second and third grid search were carried out on the best model of the first search for each feature set, to identify the optimal activation functions and learning rate (Table S3-4). Finally, a fourth grid search was carried out on the optimal model obtained from the third grid search to choose the final batch size (Table S5). In all searches, the ranking metric was the mean squared error (MSE). The optimal hyperparameters values are discussed in the next section and shown in Table S6 of the supporting information.

**RESULTS AND DISCUSSION**

**Optimal DNN models**

From our grid searches, as shown in Table 2, we found that the optimal network architecture consisted of three hidden layers with 64 neurons in the first layer, 24 in the second and 24 in the third layer.

Table 2: Optimal DNN hyperparameters found from four grid-searches with five-fold cross validation.

| Hyperparameter | Value |
| --- | --- |
| neuron configuration | (64, 24, 24) |
| epochs | 3000 |
| batch size | 128 |
| hidden layer activation function | softsign |
| output activation function | tanh |
| learning rate | 0.0005 |

The grid search on the activation function indicated the softsign function as best for the hidden layers. For the output layer, the top three functions were the softsign, linear and hyperbolic tangent functions. These three combinations had comparable metrics of loss and we chose the tanh function as the output function. Finally, the optimal learning rate was found to be 0.0005. The best batch size was determined in the fourth grid search by comparing the moving average validation loss of the optimal model and found to be 128. Plots and details of the grid search results are included in the supporting information (Figures S3-5). We would like to note that a grid search for regularization was also carried out but did not lead to an improvement on our model as overfitting was not observed.

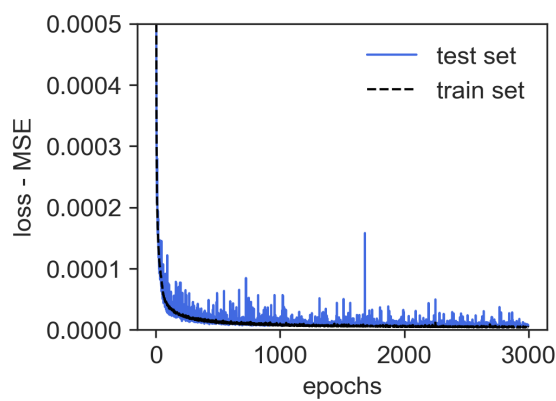

Figure 2: Final optimized DNN model training loss (MSE) (black dashed line) and testing loss (blue solid line) for the scaled logarithm of the reaction rate product $\log(k_Q)$ as a function of epochs.

In Figure 2 we show the testing dataset loss (blue solid line) and the training dataset loss (black dashed line) as a function of epochs. The testing loss closely follows the training loss and decreases as training progresses. We do not observe signs of overfitting from the trend of the average testing loss. The fluctuations in the loss are associated with the batch size in mini batch training. The final value of the training loss expressed in terms of the scaled $\log(k_Q)$ is equal to $4.02 \cdot 10^{-6}$.

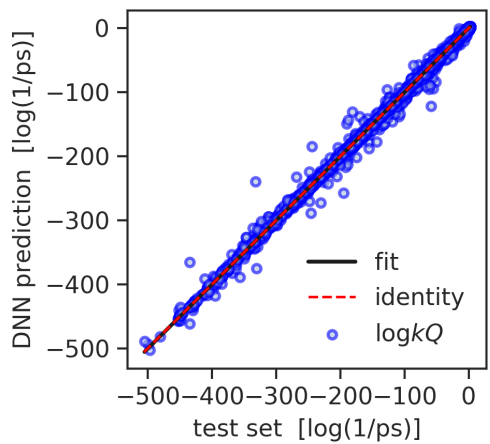

Figure 3: Predictions of the optimal DNN model on the test set. Blue circles represent the predicted value of $\log(k_Q)$ compared to the true value in the test set. A linear fit to the data (black line) is very similar to the identity function (red line). The percent MAE rescaled with respect to the mean is 1.1%. The standard deviation from the mean is 0.60 log(1/ps).

In Figure 3 we show the unscaled predicted DNN values of the $\log(k_Q)$ compared to the exact test set values (blue circles). The red dashed line indicates the identity, i.e. the case where a model would predict perfectly. The linear fit (black line) to the predicted data is very close to the identity with a mean absolute error (MAE) of 0.18 log (1/ps) and standard deviation of 0.60 log(1/ps). This corresponds to an absolute relative error of 1.1% with respect to the mean $\log(k_Q)$ value. The null hypothesis of always predicting the mean of $\log(k_Q)$ produces a MAE of 18.2 log(1/ps) with standard deviation 26.9 log(1/ps). The small value of the MAE compared to the null indicates that the model is able to predict accurately within the range of values it was trained on.

**Quantum rate constant predictions**

We investigated the trained DNN's ability to predict quantum effects by comparing predicted $\log(k_Q)$ values to transition state theory rate constant results for the entire test set.

In Figure 4 panel a) the average difference between the DNN predicted values and the corresponding transition state rate constant, binned by temperature interval, is shown in grey bars with diagonal lines. Specifically, the $j$-th bin's height $h_j$ is computed as

$$h_j = 1/N_{test}^{(j)} \cdot \sum_{i=1}^{N_{test}^{(j)}} \left( \log\left(k_{Q_i}\right) - \log\left(k_{Q_i}^{TST}\right) \right) / \left|\log\left(k_{Q_i}\right)\right| \quad (3)$$

The normalization by the absolute value of the quantum $\log(k_Q)$ is carried out so as to even out changes in rate value due to reactant mass and barrier shape and emphasize the difference with respect to the classical rate constant. The same average difference is shown in blue for the test dataset. The DNN values closely follow the test set and for temperatures below 300K, the percent MAE rescaled with respect to the exact mean difference is 31%.

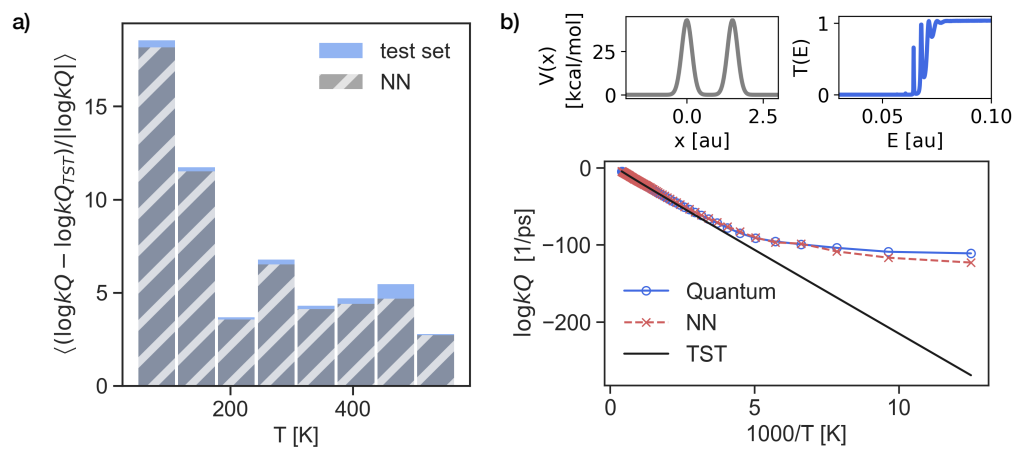

Figure 4: Panel a) average prediction of quantum effects (equation (3)) as a function of temperature for the test set (light blue) and for the neural network predictions (grey with diagonal lines). As temperature increases the NN correctly predicts a decrease in quantum effects. Panel b) Top plots show, on the left hand side, a double barrier potential taken from the test set with $V_1 = V_2 = 42.89$ kcal/mol, $w_1 = w_2 = 0.5$ au and $d = 1.5$ au and, on the right hand side, the transmission coefficient for reactants of mass 11.63 g/mol. The lower plot reports the corresponding predicted $\log(k_Q)$ as a function of the inverse temperature (red dashed line), the exact values (blue line) and a transition state theory estimate based on the first barrier height (black line).

In panel b) of Figure 4 we show an example of the predicted $\log(k_Q)$ for a symmetric double gaussian barrier taken from the test set. The barrier heights were $V_1 = V_2 = 42.89$ kcal/mol, the widths $w_1 = w_2 = 0.5$ au and the distance between barriers was $d = 1.5$ au. The reactants mass was 11.63 g/mol which is close to the mass of a carbon atom. As can be seen from the transmission coefficient, panel b) top right-hand side plot, the peaks indicate the presence of quasi bound states which lead to resonant tunneling and to a larger rate constant.[33,38] The presence of quasi bound states increases the duration of the tunneling process and contemporarily the length of time-dependent calculations required to compute the rate constant ab initio. In this context, it is promising to see that the predicted DNN $\log(k_Q)$ values (red) closely follow the exact quantum values (blue) and only start underestimating at temperatures below ~105 K.

**Predicting rate constants for symmetric single and double barriers: the H+H$_2$ reaction and hydrogen diffusion on Ni(100)**

To evaluate the ability of the DNN model to predict kinetics beyond the test dataset, two additional reactions with symmetric barriers were considered. The first is the extensively studied H + H$_2$ reaction.[39] We employed the one-dimensional Eckart barrier approximation to the potential energy surface[11] with $V_1 = 9.8$ kcal/mol, $w_1 = 2.31$ au and $m = 1060$ au to solve for the exact kinetics. The potential is shown in Figure 5 panel a) upper plot. In the lower plot, we show the exact quantum results (blue with circles), the DNN predictions (red with triangles) and transition state theory (black line). The DNN predictions follow the trend of the exact results and recovers most of the quantum effects.

The second process we studied was hydrogen diffusion on Ni(100). For this system, we extracted a one-dimensional pathway from the EDIM potential energy surface[40] with EAM-5 parametrization developed by Truong and Truhlar.[41] The potential energy path was determined by first minimizing the potential energy along the $z$ coordinate and then by taking a path symmetrically crossing three hollow sites. The one-dimensional path was then fitted to a double gaussian potential to extract features for the DNN. A plot of the two-dimensional surface and of the one-dimensional double gaussian path is shown in Figure 5, panel b) upper section. With this potential we computed the exact logarithm of the reaction rate constant product (Figure 5, panel b) lower plot – blue), the transition state theory estimate for the rate constant (black) and the DNN predicted rate constant (red).

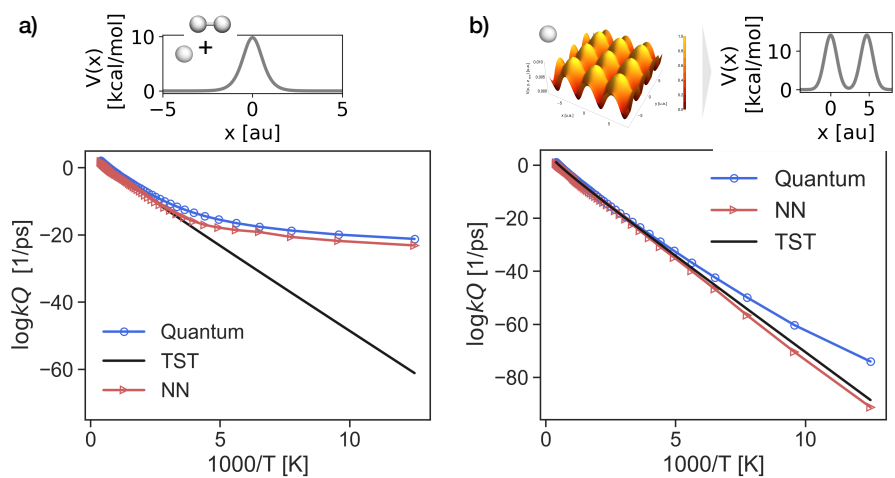

Figure 5: Panel a) Upper plot: Eckart barrier for the H + H$_2$ collinear reaction. Lower plot: comparison of the exact log ($k_Q$) for the H+H$_2$ reaction (blue with circles) with the DNN prediction (red with triangles) and transition state theory (black line). Panel b) Upper plots: two-dimensional potential energy surface for Ni(100) and extracted double gaussian fitted one-dimensional barriers. Lower plot: exact quantum log ($k_Q$) (blue with circles) for the one-dimensional diffusion of hydrogen on a Ni(100) surface compared to the DNN prediction (red with triangles) and transition state theory (black line).

The DNN predictions underestimate the rate constant at low temperatures although they are on the same order of magnitude. The error of the NN could be reduced in future work by including the TST rate constant as an input feature and training on a dataset of quantum reaction rate constants.

# Predicting rate constants for asymmetric barriers: the Menshutkin reaction, the F+HCl reaction and cyanohydrin reactions with hydrogen sulfide

We also evaluated the trained DNN's ability to predict the reaction rate constant for asymmetric barriers. These include, the Menshutkin reaction of pyridine with $CH_3Br$ in the gas phase, the reaction of F with HCl and the reaction of formalcyanohydrin with hydrogen sulfide in water. For the Menshutkin reaction, the potential[42] was fitted to an asymmetric Eckart barrier, Figure 6 top of panel a). In the bottom part of panel a) we see that the DNN predicts the rate constant very accurately. The same can also be seen in panel b) for the reaction of formalcyanohydrin with hydrogen sulfide where the potential was fitted from Ref [43]. In panel c) and d) of Figure 6 we show results for the reaction of F with HCl. The MEP was fitted to that of Ref. [44]. For panel c) the value of the activation energy, 7.21 kcal/mol, was obtained from the fit to an Eckart barrier. This value is within the range of values seen by the network. From the bottom graph we note that the DNN is able of predicting accurately. In panel d) the value of the activation energy for the same F + HCl reaction was taken from a higher level multireference coupled-cluster study[44] and is outside of the range of values seen when the network was training. This extrapolation leads to a larger error in prediction; nonetheless the DNN is accurate in predicting the rate at low temperatures.

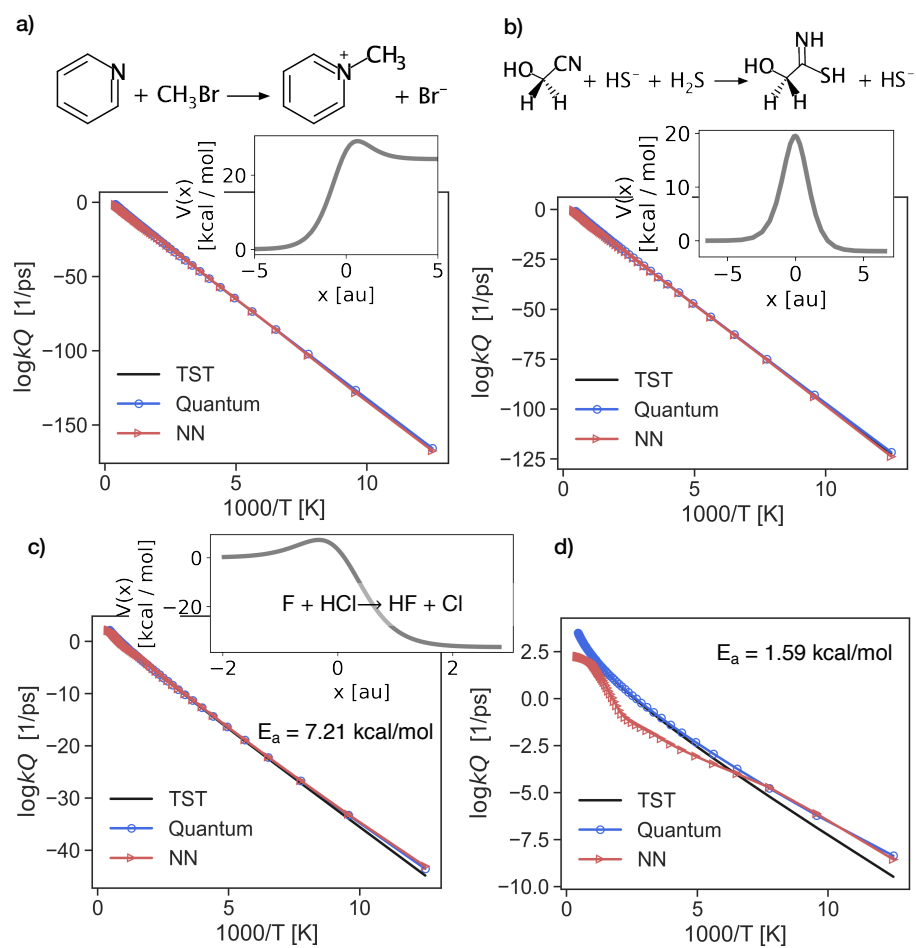

Figure 6: Panel a) Upper plot: Eckart barrier fit for the pyridine + $CH_3Br$ Menshutkin reaction. Lower plot: comparison of the exact log $(k_Q)$ for the reaction (blue with circles) with the DNN prediction (red with triangles) and transition state theory (black line). Panel b) Upper plot: Eckart barrier fit for the MEP of the reaction of formalcyanohydrin with hydrogen sulfide with an explicit solvent molecule. Lower plot: exact quantum log $(k_Q)$ (blue with circles) compared to the DNN prediction (red with triangles) and transition state theory (black line). Panels c) and d), top of panels, Eckart barrier fit to the MEP for the reaction of F with HCl. Lower plot panel c) comparison of the exact log $(k_Q)$ for the reaction (blue with circles) with the DNN prediction (red with triangles) and transition state theory (black line) when the activation energy is taken from the fit. Lower plot panel d) same as the lower plot of panel c) but with a lower more accurate activation energy.[44]

The values of the input features used to predict the rates in Figure 6 are provided in the supporting information, Table S7.

These results show that with our simple input feature representation, one can obtain qualitative information on reactivity for systems beyond the test set. To predict quantum rates more accurately and obtain quantitative accuracy for a broader set of reaction, we believe that providing information such as the reactant partition function and input features which include more information on the chemical properties of the reactants could improve the trained DNN's predictive ability.

## CONCLUSIONS

In this work we have shown that a DNN can be trained to successfully predict quantum reaction rate constant for one-dimensional reactive pathways. We generated a database of over 6.9 million quantum reaction rate constant products for smooth and rectangular, single and double, symmetric and asymmetric barriers over a broad range of temperatures and reactant masses. The database was employed to identify the optimal DNN architecture via grid search and finally to train a predictive model. The results showed a low error on predicting the test set rate constants with a relative MAE on the logarithm of $k_Q$ of 1.1%. Furthermore, at temperatures lower than 300K the trained DNN was able to predict quantum effects such as tunneling with a relative error of 31%. Beyond the testing dataset we also looked at new cases such as the H+$H_2$ reaction, the diffusion of H on Ni, the reaction of pyridine with MeBr, the reaction of F with HCl and the reaction of formalcyanohydrin with hydrogen sulfide. We found that the predicted values closely followed the exact results. We believe that these predictions could improve even further when including input features related to specific molecular reactants and their chemical properties, as well as the reactant partition function. In the future, when datasets of quantum rate constants become available, classical TST rate constants and tunneling corrections could be added as input features. While approximating the kinetics with a single one-dimensional path simplifies the multidimensional dynamics, its quantum mechanical solution remains challenging. In this context our work's trained DNN model provides a quicker alternative for the prediction of quantum rate constants and helps open the road for the long-term prospect of modeling networks of coupled reactions.

## SUPPORTING INFORMATION

The supporting information includes information on the dataset generation, the dataset features, its distribution and cross correlations. It also contains details on the grid search spaces, visualizations of the grid search optimization, and final model hyperparameters.

## ACKNOWLEDGMENTS

The authors would like the acknowledge the Hyak supercomputer system at the University of Washington for support of this research.

## REFERENCES

(1) H. Larsen, J.; Chorkendorff, I. From Fundamental Studies of Reactivity on Single Crystals to the Design of Catalysts. *Surf. Sci. Rep.* **1999**, *35* (5–8), 163–222. https://doi.org/10.1016/S0167-5729(99)00009-6.

(2) Schmidbauer, S.; Hohenleutner, A.; König, B. Chemical Degradation in Organic Light-Emitting Devices: Mechanisms and Implications for the Design of New Materials. *Adv. Mater.* **2013**, *25* (15), 2114–2129. https://doi.org/10.1002/adma.201205022.
(3) Farrauto, C. H.; Bartholomew, R. J. *Fundamentals of Industrial Catalytic Processes*; John Wiley & Sons, Ltd., 2005.
(4) Swinney, D. C. Applications of Binding Kinetics to Drug Discovery: Translation of Binding Mechanisms to Clinically Differentiated Therapeutic Responses. *Int. J. Pharm. Med.* **2008**, *22* (1), 23–34. https://doi.org/10.1007/BF03256679.
(5) Steinfeld, J. .; Francisco, J. .; Hase, W. . *Chemical Kinetics and Dynamics*; Prentice Hall Englewood Cliffs (New Jersey), 1989.
(6) Hänggi, P.; Talkner, P.; Borkovec, M. Reaction-Rate Theory: Fifty Years after Kramers. *Rev. Mod. Phys.* **1990**, *62* (2), 251–341. https://doi.org/10.1103/RevModPhys.62.251.
(7) Peters, B. *Reaction Rate Theory and Rare Events*; Elsevier Science, 2017.
(8) Truhlar, D. G.; Garrett, B. C. Variational Transition State Theory. *Ann. Rev. Phys. Chem.* **1984**, *35*, 159–189.
(9) Mandrà, S.; Valleau, S.; Ceotto, M. Deep Nuclear Resonant Tunneling Thermal Rate Constant Calculations. *Int. J. Quantum Chem.* **2013**, *113* (12), 1722–1734. https://doi.org/10.1002/qua.24395.
(10) Wigner, E. On the Quantum Correction for Thermodynamic Equilibrium. *Phys. Rev.* **1932**, *40* (5), 749–759. https://doi.org/10.1103/PhysRev.40.749.
(11) Miller, W. H.; Zhao, Y.; Ceotto, M.; Yang, S. Quantum Instanton Approximation for Thermal Rate Constants of Chemical Reactions. *J. Chem. Phys.* **2003**, *119* (3), 1329–1342. https://doi.org/10.1063/1.1580110.
(12) Yamamoto, T.; Miller, W. H. On the Efficient Path Integral Evaluation of Thermal Rate Constants within the Quantum Instanton Approximation. *J. Chem. Phys.* **2004**, *120* (7), 3086–3099. https://doi.org/10.1063/1.1641005.
(13) Miller, W. H.; Schwartz, S. D.; Tromp, J. W. Quantum Mechanical Rate Constants for Bimolecular Reactions. *J. Chem. Phys.* **1983**, *79* (10), 4889–4898. https://doi.org/10.1063/1.445581.
(14) Rupp, M.; Tkatchenko, A.; Müller, K. R.; Von Lilienfeld, O. A. Fast and Accurate Modeling of Molecular Atomization Energies with Machine Learning. *Phys. Rev. Lett.* **2012**, *108* (5), 1–5. https://doi.org/10.1103/PhysRevLett.108.058301.
(15) Montavon, G.; Rupp, M.; Gobre, V.; Vazquez-Mayagoitia, A.; Hansen, K.; Tkatchenko, A.; Müller, K. R.; Anatole Von Lilienfeld, O. Machine Learning of Molecular Electronic Properties in Chemical Compound Space. *New J. Phys.* **2013**, *15*. https://doi.org/10.1088/1367-2630/15/9/095003.
(16) Häse, F.; Valleau, S.; Pyzer-Knapp, E.; Aspuru-Guzik, A. Machine Learning Exciton Dynamics. *Chem. Sci.* **2016**, *7* (8), 5139–5147. https://doi.org/10.1039/c5sc04786b.
(17) Afzal, M. A. F.; Haghighatlari, M.; Ganesh, S. P.; Cheng, C.; Hachmann, J. Accelerated Discovery of High-Refractive-Index Polyimides via First-Principles Molecular Modeling, Virtual High-Throughput Screening, and Data Mining. *J. Phys. Chem. C* **2019**, *123* (23), 14610–14618. https://doi.org/10.1021/acs.jpcc.9b01147.
(18) Schütt, K. T.; Arbabzadah, F.; Chmiela, S.; Müller, K. R.; Tkatchenko, A. Quantum-Chemical Insights from Deep Tensor Neural Networks. *Nat. Commun.* **2017**, *8*, 6–13. https://doi.org/10.1038/ncomms13890.
(19) Winter, R.; Montanari, F.; Noé, F.; Clevert, D. A. Learning Continuous and Data-Driven Molecular Descriptors by Translating Equivalent Chemical Representations. *Chem. Sci.* **2019**, *10* (6), 1692–1701. https://doi.org/10.1039/c8sc04175j.
(20) Gómez-Bombarelli, R.; Aguilera-Iparraguirre, J.; Hirzel, T. D.; Duvenaud, D.; Maclaurin, D.; Blood-Forsythe, M. A.; Chae, H. S.; Einzinger, M.; Ha, D. G.; Wu, T.; et al. Design of Efficient Molecular Organic Light-Emitting Diodes by a High-Throughput Virtual Screening and Experimental Approach. *Nat. Mater.* **2016**, *15* (10), 1120–1127. https://doi.org/10.1038/nmat4717.
(21) Afzal, M. A. F.; Cheng, C.; Hachmann, J. Combining First-Principles and Data Modeling for the Accurate Prediction of the Refractive Index of Organic Polymers. *J. Chem. Phys.* **2018**, *148* (24). https://doi.org/10.1063/1.5007873.
(22) Amabilino, S.; Bratholm, L. A.; Bennie, S. J.; Vaucher, A. C.; Reiher, M.; Glowacki, D. R. Training Neural Nets to Learn Reactive Potential Energy Surfaces Using Interactive Quantum Chemistry in Virtual Reality. *J. Phys. Chem. A* **2019**, *123* (20), 4486–4499. https://doi.org/10.1021/acs.jpca.9b01006.
(23) Gao, H.; Struble, T. J.; Coley, C. W.; Wang, Y.; Green, W. H.; Jensen, K. F. Using Machine Learning to Predict Suitable Conditions for Organic Reactions. *ACS Cent. Sci.* **2018**, *4* (11), 1465–1476. https://doi.org/10.1021/acscentsci.8b00357.
(24) Singh, A. R.; Rohr, B. A.; Gauthier, J. A.; Nørskov, J. K. Predicting Chemical Reaction Barriers with a Machine Learning Model. *Catal. Letters* **2019**, *149* (9), 2347–2354. https://doi.org/10.1007/s10562-019-02705-x.
(25) Ulissi, Z. W.; Medford, A. J.; Bligaard, T.; Nørskov, J. K. To Address Surface Reaction Network Complexity Using Scaling Relations Machine Learning and DFT Calculations. *Nat. Commun.* **2017**, *8*. https://doi.org/10.1038/ncomms14621.
(26) Houston, P. L.; Nandi, A.; Bowman, J. M. A Machine Learning Approach for Prediction of Rate Constants. *J. Phys. Chem. Lett.* **2019**, *10* (17), 5250–5258. https://doi.org/10.1021/acs.jpclett.9b01810.
(27) Grambow, C. A.; Pattanaik, L.; Green, W. H. Deep Learning of Activation Energies. *J. Phys. Chem. Lett.* **2020**, *11* (8), 2992–2997. https://doi.org/10.1021/acs.jpclett.0c00500.

(28) Bhoorasingh, P. L.; Slakman, B. L.; Seyedzadeh Khanshan, F.; Cain, J. Y.; West, R. H. Automated Transition State Theory Calculations for High-Throughput Kinetics. *J. Phys. Chem. A* **2017**, *121* (37), 6896–6904. https://doi.org/10.1021/acs.jpca.7b07361.

(29) Lawson, A. J.; Swienty-Busch, J.; Géoui, T.; Evans, D. The Making of Reaxys - Towards Unobstructed Access to Relevant Chemistry Information. In *ACS Symposium Series*; 2014. https://doi.org/10.1021/bk-2014-1164.ch008.

(30) SciFinder Scholar. *Choice Rev. Online* **2007**. https://doi.org/10.5860/choice.44-6577.

(31) Tromp, J. W.; Miller, W. H. The Reactive Flux Correlation Function for Collinear Reactions H + H2, Cl + HCl and F + H2. *Faraday Discuss. Chem. Soc.* **1987**, *84*, 441–453. https://doi.org/10.1039/DC9878400441.

(32) Eckart, C. The Penetration of a Potential Barrier by Electrons. *Phys. Rev.* **1930**, *35* (11), 1303–1309. https://doi.org/10.1103/PhysRev.35.1303.

(33) Berman, L.; Peskin, U. Resonant Tunneling Probabilities for an N-Terminal Junction by the Flux Averaging Method. In *International Journal of Quantum Chemistry*; 2004. https://doi.org/10.1002/qua.20024.

(34) Ahmed, Z. Tunneling through a One-Dimensional Potential Barrier. *Phys. Rev. A* **1993**, *47* (6), 4761–4767.

(35) Pedregosa, F.; Varoquaux, G.; Gramfort, A.; Michel, V.; Thirion, B.; Grisel, O.; Blondel, M.; Prettenhofer, P.; Weiss, R.; Dubourg, V.; et al. Scikit-Learn: Machine Learning in Python. *J. Mach. Learn. Res.* **2011**.

(36) Chollot, F. and others. Keras. *GitHub* **2015**.

(37) Kingma, D. P.; Ba, J. L. Adam: A Method for Stochastic Optimization. *3rd Int. Conf. Learn. Represent. ICLR 2015 - Conf. Track Proc.* **2015**, 1–15.

(38) Yamamoto, H. Resonant Tunneling Condition and Transmission Coefficient in a Symmetrical One-Dimensional Rectangular Double-Barrier System. *Appl. Phys. A Solids Surfaces* **1987**, *42* (3), 245–248. https://doi.org/10.1007/BF00620608.

(39) Truhlar, D. G.; Wyatt, R. E. History of H3 Kinetics. *Annu. Rev. Phys. Chem.* **1976**, *27* (1), 1–43. https://doi.org/10.1146/annurev.pc.27.100176.000245.

(40) Truong, T. N.; Truhlar, D. G.; Garrett, B. C. Embedded Diatomics-in-Molecules: A Method to Include Delocalized Electronic Interactions in the Treatment of Covalent Chemical Reactions at Metal Surfaces. *J. Phys. Chem.* **1989**, *93* (25), 8227–8239. https://doi.org/10.1021/j100362a017.

(41) Wonchoba, S. E.; Hu, W. P.; Truhlar, D. G. Surface Diffusion of H on Ni(100): Interpretation of the Transition Temperature. *Phys. Rev. B* **1995**, *51* (15), 9985–10002. https://doi.org/10.1103/PhysRevB.51.9985.

(42) Castejon, H.; Wiberg, K. B. Solvent Effects on Methyl Transfer Reactions. 1. The Menshutkin Reaction. *J. Am. Chem. Soc.* **1999**, *121* (10), 2139–2146. https://doi.org/10.1021/ja983736t.

(43) Valleau, S.; Martínez, T. J. Reaction Dynamics of Cyanohydrins with Hydrosulfide in Water. *J. Phys. Chem. A* **2019**, *123* (33). https://doi.org/10.1021/acs.jpca.9b05735.

(44) Aoto, Y. A.; Köhn, A. Revisiting the F + HCl → HF + Cl Reaction Using a Multireference Coupled-Cluster Method. *Phys. Chem. Chem. Phys.* **2016**, *18* (44), 30241–30253. https://doi.org/10.1039/c6cp05782a.

# Supporting Information for

# "Machine Learning Quantum Reaction Rate Constants"

Evan Komp[*,1] and Stéphanie Valleau,[*,2]

[*]Department of Chemical Engineering, University of Washington, Seattle, Washington 98115, United States

## Table of Contents

---

[1] email: evankomp@uw.edu

[2] email: valleau@uw.edu

# I. DATABASE GENERATION

The equations for the potentials employed to generate the database are defined below in Table S1. For the single barriers the exact expressions for the transmission coefficient $T(E)$ were used.[1,2] The first asymmetry parameter $\alpha$ for a single barrier was defined as the difference in energy between products and reactants. The second asymmetry parameter $\alpha_2$ was defined as $\alpha_2 = \frac{V_1-V_2}{V_1+V_2} + \frac{w_1-w_2}{w_1+w_2}$, with $V_1, V_2$ the height of the barriers and $w_1, w_2$ their width.

For double barriers, the transmission coefficient was evaluated as in equation (2) of the main text[3], with $x \in [-L, L]$ and $L = 20$ [au]. The evaluation of the transmission coefficient was carried out using scipy's `integrate.solve_ivp` function to solve the initial value problem[4] with the implicit Runge-Kutta method of order 5 as defined by the 'Radau' option. The range of energy values over which the transmission coefficient was evaluated was a set of 300 points in the interval $E_{values} = \left[\frac{E_{max}}{1000}; 2E_{max}\right]$, with $E_{max} = 2 \cdot \max[V(x)]$. The maximum value of the potential energy was evaluated over 1000 points in the range $-L < x < L$.

Table S1: Equations for the one-dimensional potential energy barriers used to build the database.

| Barrier | Potential energy | Parameters |
|---|---|---|
| Single rectangular | $V_R(x) = \begin{cases} V_1 & 0.0 < x < w_1 \\ 0.0 & x < 0 \lor x > w_1 \end{cases}$ | $V_1$ – barrier height<br>$w_1$ – barrier width<br>$\alpha_1 = 0.0$<br>$\alpha_2 = 0.0$<br>$s = 0.0$ |
| Single symmetric Eckart | $V_{SE}(x) = \frac{V_1}{\cosh^2(\pi x/w_1)}$ | $V_1$ – barrier height<br>$w_1$ – barrier width<br>$\alpha_1 = 0.0$<br>$\alpha_2 = 0.0$<br>$s = \frac{V(x^*) - V\left(x^* + \frac{w_1}{2} - \Delta x\right)}{w_1/2}$<br>$x^* \mid V(x^*) = \max(V(x)))$<br>$\Delta x = 0.025$ [au] |
| Single asymmetric Eckart | $V_{AE}(x) = \frac{V_1(1-\alpha)}{1+e^{-2\pi x/w_1}} + \frac{V_1(1+\sqrt{\alpha})^2}{4\cosh^2 \pi x/w_1}$ | $V_1$ – barrier height<br>$w_1$ – barrier width<br>$\alpha_1 = V_{react} - V_{prod}$<br>$\alpha_2 = 0.0$<br>$s = \frac{\|V(x^*) - V(x_+)\| + \|V(x^*) - V(x_-)\|}{w_1}$<br>$x^* \mid V(x^*) = \max(V(x))$<br>$x_\pm = x^* \pm \frac{w_1}{2} \mp \Delta x \ ; \ \Delta x = 0.025$ [au] |
| Double rectangular | $V_{DR}(x) = \begin{cases} V_1 & 0.0 < x < w_1 \\ V_2 & w_1 + \Delta x_{12} < x < w_1 + \Delta x_{12} + w_2 \\ 0.0 & \text{elsewhere} \end{cases}$ | $V_i$ – barrier *i* height<br>$w_i$ – barrier *i* width<br>$d = \Delta x_{12} + (w_1 + w_2)/2$<br>$\alpha_1 = 0$<br>$\alpha_2 = \frac{V_1 - V_2}{V_1 + V_2} + \frac{w_1 - w_2}{w_1 + w_2}$<br>$s = 0.0$ |
| Double gaussian | $V_{DG}(x) = V_1 e^{\frac{-(x-x_1)^2}{2\sigma_1^2}} + V_2 e^{\frac{-(x-x_2)^2}{2\sigma_2^2}}$<br>$\sigma_1 = w_1/3 ; \quad \sigma_2 = w_2/3$<br>$x_1 = 0.0 ; \quad x_2 = 3 \cdot (\sigma_1 + \sigma_2) + d$ | $V_i$ – barrier *i* height<br>$w_i$ – barrier *i* width<br>$d$ – distance between barriers<br>$\alpha_1 = 0$<br>$\alpha_2 = \frac{V_1 - V_2}{V_1 + V_2} + \frac{w_1 - w_2}{w_1 + w_2}$<br>$s = (s_1 + s_2)/2$<br>$s_i = \frac{V(x_i^*) - V\left(x_i^* + \frac{w_i}{2} - \Delta x\right)}{\frac{w_i}{2}}$<br>$x_i^* \mid V(x_i^*) = \max(V(x)) ; i = 1,2$<br>$\Delta x = 0.025 \, [au]$ |
| Peskin barrier | $V_{Pesk}(x) = V'\left(\frac{1}{\cosh^2(x)} - \frac{1}{\cosh^2(\gamma x)}\right)$ | $V'$ – constant proportional to barrier height<br>$\gamma$ – constant proportional to barrier width<br>$w_1 = w_2 = \|x^* - x_{min}\|$ barrier width<br>$\alpha_1 = 0.0$<br>$\alpha_2 = 0.0$<br>$s = (s_1 + s_2)/2 s_i = \frac{V(x_i^*) - V\left(x_i^* + \frac{w_i}{2} - \Delta x\right)}{w_i/2}$ ; difference taken with respect to $x = 0.0$ when $V_{max} > 0 \wedge V_{min} > 0$<br>$x^* \mid V(x^*) = \max(V(x))$ |

## II. DATASET

A visualization of the distribution of the training and testing dataset with respect to its input features and output label is shown in Figures S1a-c. The first plot (a) displays scattering of pairwise feature distributions with kernel density estimate (KDE) diagonals for the entire dataset. The following plots (b-c) show the pairwise distribution of subsets of features and labels for the test and train sets through KDE. Figure S2 maps the correlation between variables through the Pearson correlation coefficient.

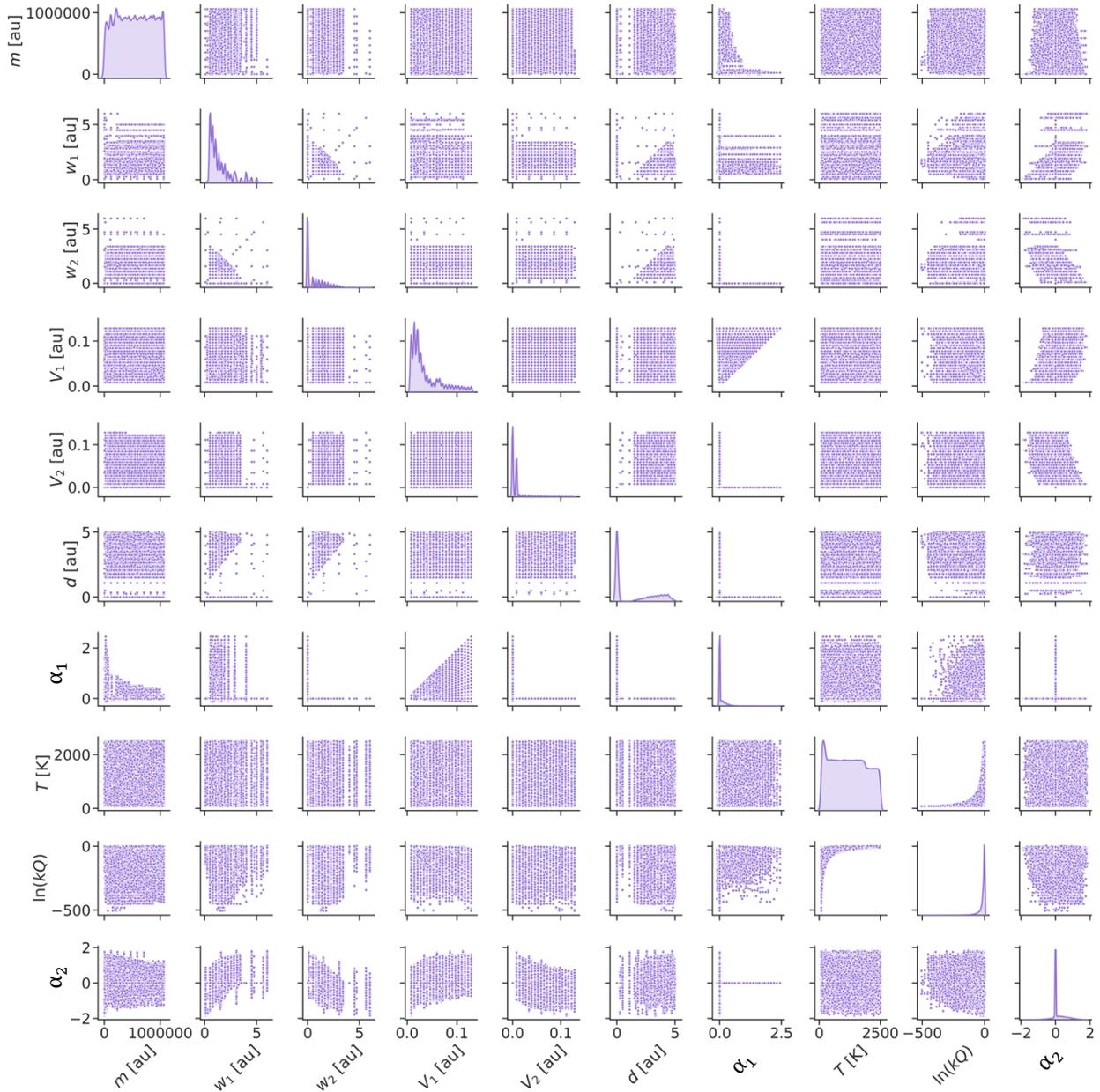

Figure S1a: Database pair-wise distribution for all input features and for the output label. Points are datum. Diagonals are gaussian KDE of single features.

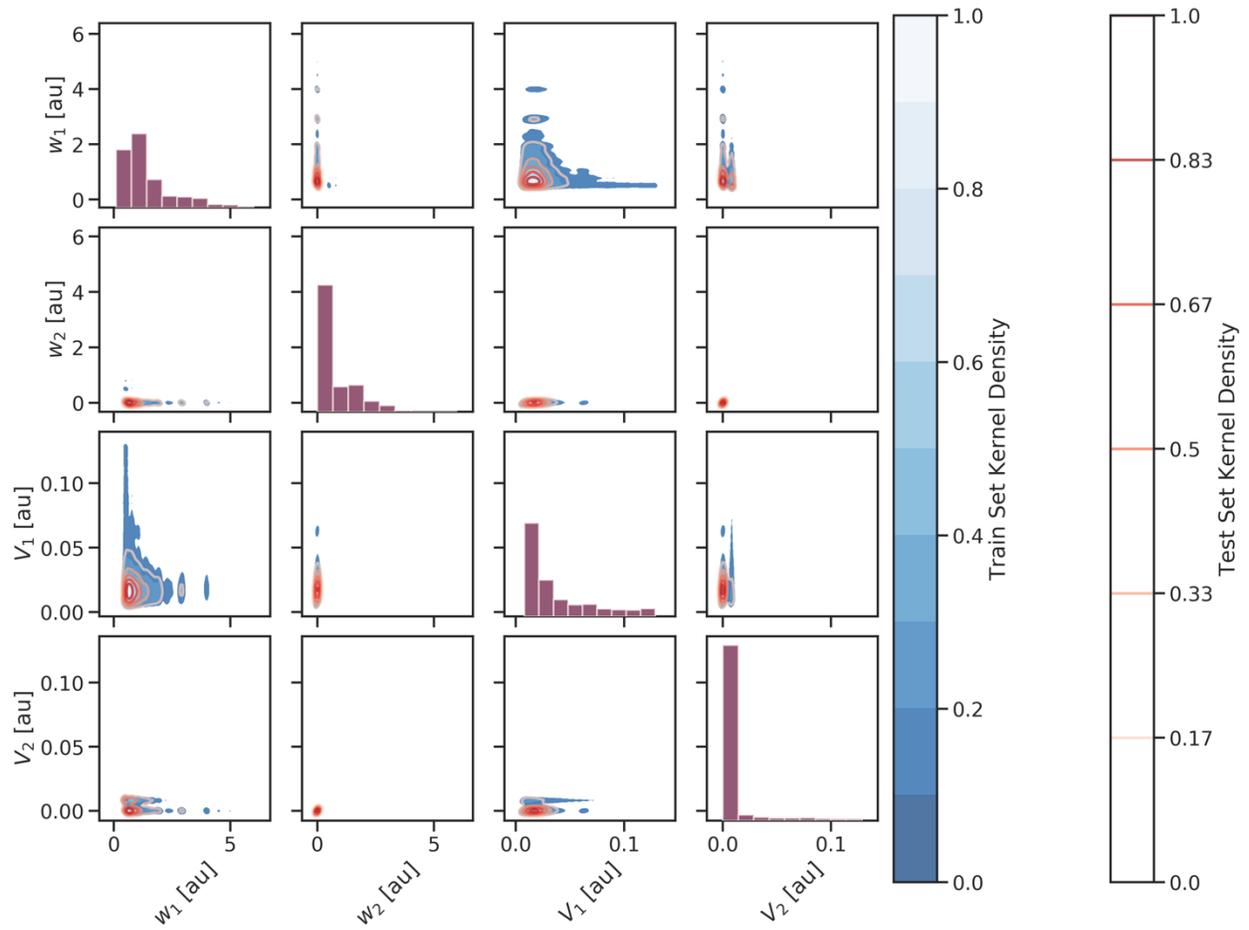

Figure S1b: Pair-wise distribution of a subset of the database features related to the barrier for the train and test sets. Blue surfaces are train set KDE, while red lines are test set KDE. KDE is normalized to 1.0. A KDE of 1.0 then indicates the region of two-feature space with the highest density of data.

When looking at the diagonal distribution of weights and heights in Figure S1b, we see a large number of points in the first bin. This comes from the fact that for single barriers the distance between barriers, the second barrier width and height are all equal to zero.

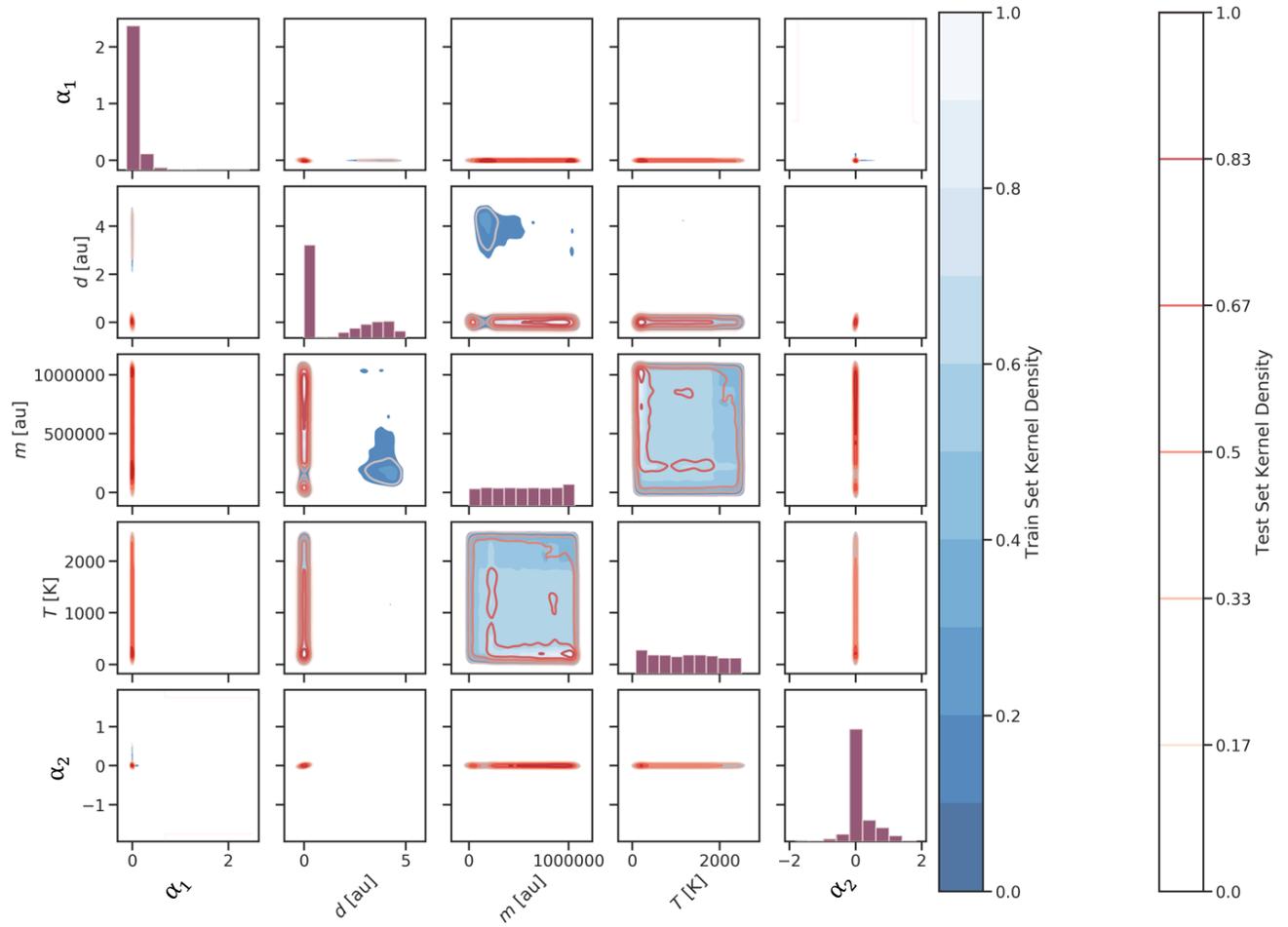

Figure S1c: Pair-wise distribution of a subset of the database features for the train and test sets. Blue surfaces are train set KDE, while red lines are test set KDE. KDE is normalized to 1.0. A KDE of 1.0 then indicates the region of two-feature space with the highest density of data.

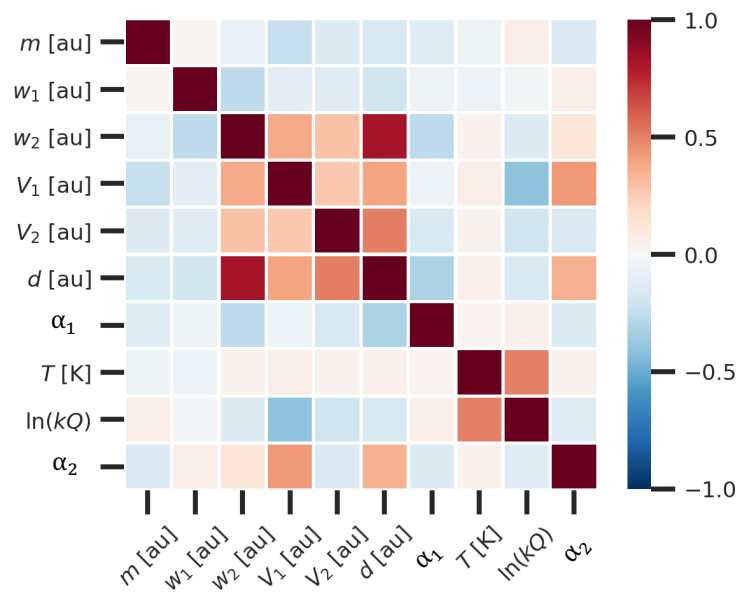

Figure S2: Correlation matrix of the Pearson correlation coefficient for input features of the database. Values close to 1.0 indicate a strong positive correlation between variables, while -1.0 indicates a strong negative correlation. Features are perfectly correlated to themselves, as can be seen by the dark red diagonal. Most features have a small correlation, close to zero which indicates little bias. One distinct correlation is between barrier distance and second barrier width. This is due to the fact that for single barriers $d$ and $w_2$ are zero while for double barriers both are non-zero. We also note a positive correlation between temperature and rate, which comes from the physics of reactivity.

## III. GRID SEARCH & HYPERPARAMETER TUNING

As described in the main text, four subsequent grid searches were carried out. The hyperparameter search spaces for each are shown in Table S2-5, and optimal hyperparameters are shown in Table S6. Figure S 3 indicates how the logarithm of the mean absolute error, MSE, averaged over all runs, changes with the number of neurons for each single hidden layer. Figure S 4 shows how the loss changes with the learning rate. Figure S5 shows validation loss and moving average validation loss during training for a range of batch sizes, highlighting the optima.

### A. First grid search: neurons, hidden layers, batch size and epochs

For the first grid search we searched over number of hidden layers, number of neurons per layer, and batch size. The range of parameters is shown in Table S2 below and the results in the following Figure S 3. We found that the optimal model was of neuron configuration (64, 24, 24).

Table S2: First grid search space values for number of hidden layers, number of neurons per layer, batch size and epochs.

| Hyperparameter | Search space |
|---|---|
| number of hidden layers | 1, 2, 3 |
| number of neurons in hidden layer | 1, 6, 12, 24, 32, 64, 128 |
| batch size | 32, 64, 128 |
| **Fixed hyperparameters** | **Default values** |
| hidden layer activation function | softsign |
| output activation function | tanh |
| epochs | 300 |
| optimizer | Adam |
| - learning rate | 0.001 |
| - beta 1 | 0.9 |
| - beta 2 | 0.9999 |
| weight initialization | random uniform |
| - range | [0, 1] |
| loss function | mean squared error (MSE) |
| **Optimal hyperparameters** | **Values** |
| neuron configuration | (64, 24, 24) |
| batch size | 64 |

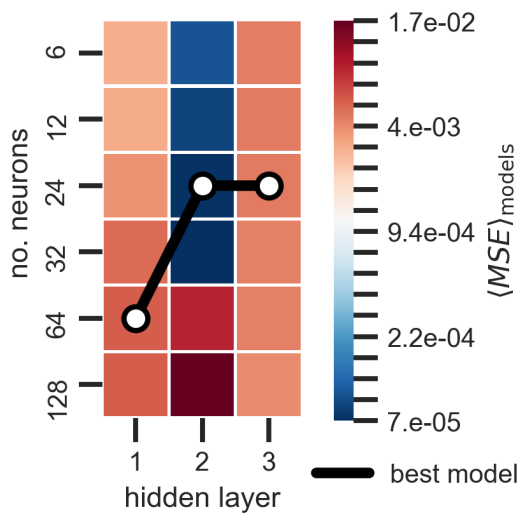

Figure S 3: Average model validation mean square error (MSE) associated with number of neurons in each of three hidden layers. The color scale corresponds to the average of the MSE over all models with three hidden layers trained for 300 epochs with batch size 64. The black line shows the neuron configuration of the optimal model.

## B.  Second grid search: activation function

For the second grid search, the choice of activation functions applied to each neuron in the hidden and output layers was explored. The activation functions tested are seen below in Table S3. The optimum configuration was found to be softsign, tanh for hidden and output layers, respectively.

Table S3: Second grid search space values for activation function in both hidden and output layers.

| Hyperparameter | Search space |
| --- | --- |
| activation function | softmax, softplus, softsign, relu, tanh, sigmoid, hard sigmoid linear |
| **Fixed hyperparameters** | **Default values** |
| neuron configurations | (64, 24, 64) |
| epochs | 300 |
| batch size | 32 |
| optimizer | Adam |
| -   learning rate | 0.001 |
| -   beta 1 | 0.9 |
| -   beta 2 | 0.9999 |
| weight initialization | random uniform |
| -   range | [0, 1] |
| loss function | mean squared error (MSE) |
| **Optimal hyperparameters** | **Values** |
| hidden layer activation | softsign |
| output layer activation | tanh |

## C.  Third grid search: learning rate

The third grid search optimized the value of learning rate for the Adam optimizer. The range of values explored were on the logarithmic scale to best search hyperparameter space and are shown below in Table S4. This was conducted on a randomly selected portion of 30% of the training set as well as on 100% of the training set. In both cases, the best value was 0.0005. The results are also shown in Figure S 4.

Table S4: Third grid search space values for the learning rate.

| Hyperparameter | Search space |
|---|---|
| learning rate | 1E-5, 5E-5, 1E-4, 5E-4,1E-3, 5E-3,1E-2, 5E-2, 1E-1, 2E-1 |
| **Fixed Hyperparameters** | **Default values** |
| neuron configuration | (64, 24, 24) |
| epochs | 300 |
| batch size | 64 |
| hidden layer activation function | softsign |
| output activation function | tanh |
| optimizer | Adam |
| -    beta 1 | 0.9 |
| -    beta 2 | 0.9999 |
| weight initialization | random uniform |
| -    range | [0, 1] |
| loss function | mean squared error (MSE) |
| **Optimal hyperparameters** | **Values** |
| learning rate | 0.0005 |

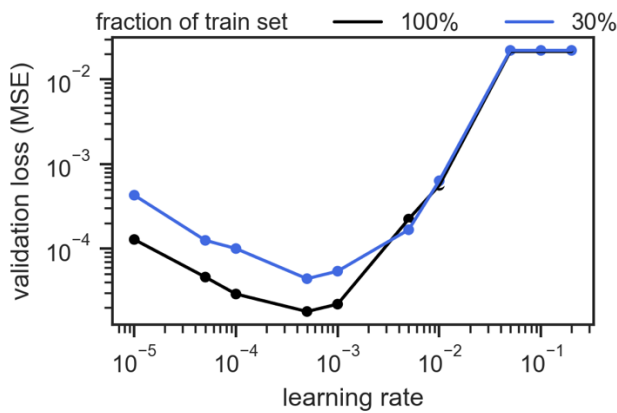

Figure S 4: Validation loss as a function of learning rate in the third grid search. The blue and black lines correspond to grid search on 30% and 100% of the training dataset. Both searches indicate that 0.0005 is the optimal learning rate value.

## D. Fourth grid search: batch size for optimal model

A final grid search on the batch size was conducted to evaluate how fluctuations on the validation loss related to mini batch training batch sizes. Details of the search are described in Table S5. A moving average of the validation loss was used as a metric, to average the impact of the fluctuations. It was found that 128, and 256 performed best in terms of reduction in fluctuation and final validation loss magnitude and given that 128 had a lower final rolling average loss it was chosen as the best batch size. This result is shown in Figure S5.

Table S5: Fourth grid search space for batch size.

| Hyperparameter | Search space |
| --- | --- |
| batch size | 32, 64, 128, 256, 512, 1024, 2048, 4096, 8192 |
| **Fixed Hyperparameters** | **Default values** |
| neuron configuration | (64, 24, 24) |
| epochs | 300 |
| hidden layer activation function | softsign |
| output activation function | tanh |
| optimizer | Adam |
| - learning rate | 0.0005 |
| - beta 1 | 0.9 |
| - beta 2 | 0.9999 |
| weight initialization | random uniform |
| - range | [0, 1] |
| loss function | mean squared error (MSE) |
| **Optimal hyperparameters** | **Values** |
| batch size | 128 |

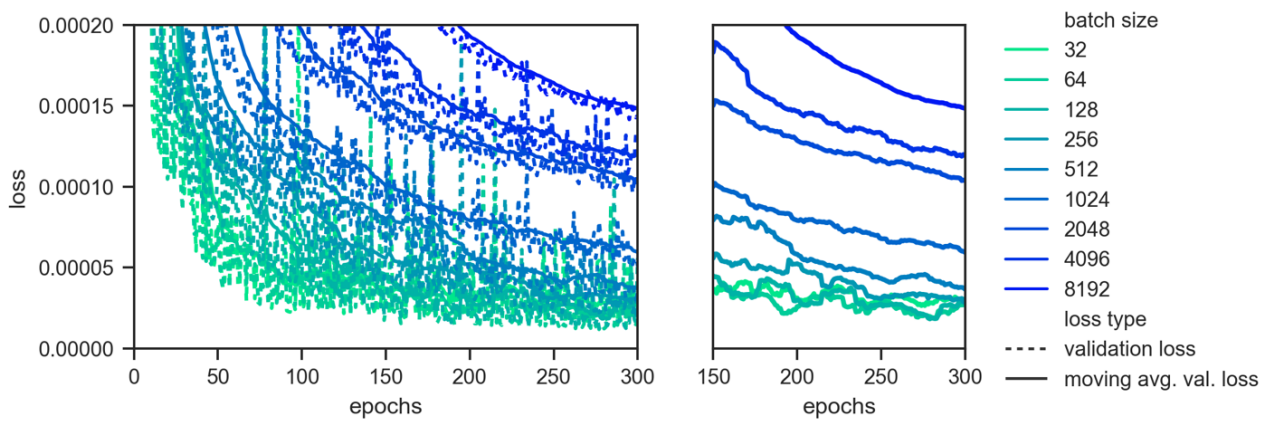

Figure S5: Left panel) Validation loss and moving average of validation loss over 20 points for models trained with a range of batch sizes as a function of epochs. Batch size is seen to impact both the fluctuations in the validation loss, as well as the final training and validation loss value. Right panel) Moving average validation loss for the last 100 epochs of training. Batch sizes 128, and 256 perform best.

## E. Optimal hyperparameters

Table S6: Optimal hyperparameters identified from grid searches for the final training.

| Hyperparameter | Value |
|---|---|
| neuron configuration | (64, 24, 24) |
| epochs | 300 |
| batch size | 128 |
| hidden layer activation function | softsign |
| output activation function | tanh |
| optimizer | Adam |
| - learning rate | 0.0005 |
| - beta 1 | 0.9 |
| - beta 2 | 0.9999 |
| weight initialization | random uniform |
| - range | [0, 1] |
| loss function | mean squared error (MSE) |
| **Performance** | |
| training loss | 4.02E-6 |
| test loss | 6.18E-6 |
| test MAE [log(1/ps)] | 0.18 (0.60) |

## IV. INPUT FEATURES FOR RATE CONSTANTS PREDICTED IN FIGURE 6 OF MAIN TEXT

Table S7 contains the values of the input features provided to the trained DNN to compute reaction rate constants for the reactions shown in Figure 6 of the main text. The input features were obtained by fitting existing minimum energy paths to asymmetric Eckart barriers.

Table S7: Input feature values used to make predictions on the rate product for asymmetric reactions outside of the testing set. Reaction 1 is the Menshutkin reaction of pyridine with methyl bromide in the gas phase.[5] Reaction 2 is the reaction of formalcyanohydrin with hydrogen sulfide in the liquid phase,[6] and Reaction 3 is the reaction of F with HCl.[7] Reaction 3a uses the activation energy obtained from the Eckart fit of the minimum energy path, while Reaction 3b uses a higher level of theory activation energy.[7]

| Feature | Reaction 1 | Reaction 2 | Reaction 3a | Reaction 3b |
|---|---|---|---|---|
| mass [au] | 317257.3 | 226395.5 | 101090.8 | 101090.8 |
| $V_1$ [au] | 0.04223043 | 0.03107522 | 0.01149779 | 0.002533826 |
| $V_2$ [au] | 0.0 | 0.0 | 0.0 | 0.0 |
| slope [au] | 0.008847832 | 0.01289512 | 0.03025426 | 0.03025426 |
| $w_1$ [au] | 4.314845 | 4.0 | 2.303736 | 2.303736 |
| $w_2$ [au] | 0.0 | 0.0 | 0.0 | 0.0 |
| d [au] | 0.0 | 0.0 | 0.0 | 0.0 |
| $\alpha_1$ | -0.03879398 | 0.003222610 | 0.05662957 | 0.05662957 |
| $\alpha_2$ | 0.0 | 0.0 | 0.0 | 0.0 |

## REFERENCES

(1) Eckart, C. The Penetration of a Potential Barrier by Electrons. *Phys. Rev.* **1930**, *35* (11), 1303–1309. https://doi.org/10.1103/PhysRev.35.1303.

(2) Ahmed, Z. Tunneling through a One-Dimensional Potential Barrier. *Phys. Rev. A* **1993**, *47* (6), 4761–4767. https://doi.org/10.1103/PhysRevA.47.4761.

(3) Mandrà, S.; Valleau, S.; Ceotto, M. Deep Nuclear Resonant Tunneling Thermal Rate Constant Calculations. *Int. J. Quantum Chem.* **2013**, *113* (12), 1722–1734. https://doi.org/10.1002/qua.24395.

(4) Virtanen, P.; Gommers, R.; Oliphant, T. E.; Haberland, M.; Reddy, T.; Cournapeau, D.; Burovski, E.; Peterson, P.; Weckesser, W.; Bright, J.; et al. SciPy 1.0: Fundamental Algorithms for Scientific Computing in Python. *Nat. Methods* **2020**, *17*, 261–272. https://doi.org/https://doi.org/10.1038/s41592-019-0686-2.

(5) Castejon, H.; Wiberg, K. B. Solvent Effects on Methyl Transfer Reactions. 1. The Menshutkin Reaction. *J. Am. Chem. Soc.* **1999**, *121* (10), 2139–2146. https://doi.org/10.1021/ja983736t.

(6) Valleau, S.; Martínez, T. J. Reaction Dynamics of Cyanohydrins with Hydrosulfide in Water. *J. Phys. Chem. A* **2019**, *123* (33), 7210–7217. https://doi.org/10.1021/acs.jpca.9b05735.

(7) Aoto, Y. A.; Köhn, A. Revisiting the F + HCl → HF + Cl Reaction Using a Multireference Coupled-Cluster Method. *Phys. Chem. Chem. Phys.* **2016**, *18* (44), 30241–30253. https://doi.org/10.1039/c6cp05782a.